\UseRawInputEncoding
\documentclass[authoryear,12pt]{elsarticle} 
\usepackage[english]{babel}
\usepackage{amsmath}
\usepackage[toc,page]{appendix}
\usepackage{amssymb}       
\usepackage{amsthm}       
\usepackage{natbib}        
\usepackage{graphicx}
\usepackage{booktabs}
\usepackage{subfigure}      
\usepackage{color}
\usepackage{gensymb} 

\journal{Journal of the Mechanics and Physics of Solids, accepted}

\newcommand{\tensor}[1]{\boldsymbol{#1}}
\newcommand{\ftensor}[1]{\mathbb{#1}}

\newcommand{\rd}{\mathrm{d}}

\begin{document}

\begin{frontmatter}

\title{Dislocation dynamics prediction of the strength of Al-Cu alloys containing shearable $\theta''$ precipitates} 

\author{R. Santos-G{\"u}emes$^{1, 2}$}
\author{L. Capolungo$^{3}$}
\author{J. Segurado$^{1, 2}$}
\author{J. LLorca$^{1, 2, }$\corref{cor1}}
\address{$^1$ IMDEA Materials Institute, C/ Eric Kandel 2, 28906, Getafe, Madrid, Spain. \\  $^2$ Department of Materials Science, Polytechnic University of Madrid/Universidad Polit\'ecnica de Madrid, E. T. S. de Ingenieros de Caminos. 28040 - Madrid, Spain. \\  $^3$ Material Science and Technology Division, MST-8, Los Alamos National Laboratory, Los Alamos 87545 NM, USA.}

\cortext[cor1]{Corresponding author}

\begin{abstract}

The critical resolved shear stress of an Al 4 wt. \% Cu alloy containing a homogeneous distribution of $\theta''$ precipitates was determined by means of dislocation dynamics simulations. The size distribution, shape, orientation and volume fraction of the precipitates in the alloy were obtained from transmission electron microscopy observations while the parameters  controlling the dislocation/precipitate interactions (elastic mismatch, transformation strains, dislocation mobility and cross-slip probability, etc.) were calculated from atomistic simulations. The precipitates were assumed to be either impenetrable or shearable by the dislocations, the latter characterized by a threshold shear stress that has to be overcome to shear the precipitate. The predictions of the simulations in terms of the critical resolved shear stress and of the dislocation/precipitate interaction mechanisms were in good agreement with the experimental results. It was concluded that the optimum strength of this alloy is attained with a homogeneous distribution of $\theta''$ precipitates whose average size ($\approx$ 40 nm) is at the transition between precipitate shearing and looping. Overall, the dislocation dynamics strategy presented in this paper is able to provide quantitative predictions of precipitate strengthening in metallic alloys.

\end{abstract}


\end{frontmatter}

\section{Introduction}
\label{sec:intro}

Metallic alloys can be strengthened by different mechanisms \citep{argon2008strengthening,BalakrishnaBhat1980} and precipitate hardening is one of the most effective among them \citep{ardell1985precipitation,Martin1998_PrecipitationHardening}. This mechanism is triggered by the presence of second phases or precipitates in the crystalline solid that hinder the movement of the dislocations in the slip planes. Therefore, the resolved shear stress on the slip plane has to be increased to overcome the obstacles, enhancing the overall strength of the material. The underlying physics of the dislocation-precipitate interaction is very complex and depend on many factors (chemical composition and crystal lattice of matrix and precipitate, size, shape and spatial distribution of the precipitates, interface coherency, etc.) but the simplest classification depends on whether the precipitates can be sheared or not by dislocations.

The development of precipitate hardening models started with relatively simple approaches based on the line tension concept. The case of impenetrable precipitates, that cannot be sheared by dislocations, was analyzed by  \cite{Orowan1948}. He considered a regular arrangement of spherical precipitates that were circumvented by the dislocation, leaving a closed loop around the precipitate, known as Orowan loop. This model was subsequently refined to take into account other phenomena such as a random spatial distribution of precipitates \citep{Kocks1966}, the interaction between dislocation arms \citep{Bacon1973_BKS} or different precipitate geometries \citep{nie1996effect}. Shearable precipitates were included in line tension models introducing a critical shear stress at which the dislocation was able to penetrate the precipitate. The critical shear stress, that determined the strength of the obstacle from the viewpoint of dislocation shearing,  could be determined from the angle between the dislocation arms when the dislocation overcomes the precipitate \citep{Friedel64,Martin1998_PrecipitationHardening}. The first computer simulations of precipitation hardening were based on these line tension models and analyzed the propagation of the dislocation line in a forest of precipitates which were represented by point obstacles in the slip plane \citep{Foreman1966_FM, Nogaret2006}.

In general, line tension models were able to provide qualitative explanations of the effect of a number of factors (i.e precipitate size, shape, spatial distribution) on the strength of precipitation-hardened alloys. However,  they could not provided quantitative predictions because many physical mechanisms affecting the dislocation/precipitate interaction were not included. These could be taken into account, however, within the framework of discrete dislocation dynamics (DDD) simulations, which were used to simulate dislocation/precipitate interactions in the case of both  shearable \citep{Mohles1999_DD,Mohles2001} and impenetrable precipitates \citep{Monnet2006_DDD_precs,Monnet2011_DDD_precs}. Both types of precipitates were also studied using the level-set method \citep{Xiang2004_LevelSet,Xiang2006_LevelSet}, which also included the effect of the misfit dilatational strain, and pointed out the relevance of out-of-plane dislocation movements, such as cross-slip or climb, to overcome the precipitates. Other investigations considered the effect of the elastic heterogeneity between the precipitate and the matrix \citep{Shin2003_DDD_FEM,Takahashi2008_DDD_BEM,Takahashi2011_DDD_FEM}, which was found to be small in most cases, and of dislocation cross-slip \citep{Shin2003_DDD_FEM, Monnet2006_DDD_precs}. Nevertheless, all these simulations were limited to ideal cases in which the shape of the precipitates was spherical and they were arranged in regular arrays. As a result, they could be used to enrich line tension models but still lacked the capability to make quantitative predictions of realistic precipitate distributions in actual alloys.

More recently, \cite{Santos-Guemes2018}  presented a DDD framework to study dislocation/precipitate interactions based on the discrete-continuous approach, originally developed by \cite{Lemarchand2001_DCM}. This strategy was used by \cite{Santos-Guemes2020} to simulate precipitation hardening in the case of an Al-Cu alloy containing a homogeneous dispersion of $\theta'$ precipitates that cannot be sheared by dislocations. DDD simulations were carried out in a  1 x 1 x 1 $\mu$m$^3$ domain which included 12 precipitates, whose size, shape, crystallographic orientation and volume fraction were obtained from experimental observations. Moreover, the influence of solution hardening, transformation strains, matrix-precipitate elastic mismatch and dislocation cross-slip was taken into account in the simulations. The critical resolved shear stress (CRSS) obtained by averaging 12 simulations was in excellent agreement with the experimental values given by micropillar compression tests of single crystals, demonstrating the ability of the model to make quantitative predictions of precipitate strengthening. 

In this investigation, the approach used in \cite{Santos-Guemes2020} is extended to include precipitate shearing and it is applied to determine the CRSS of a peak-aged Al-Cu alloy containing a homogeneous dispersion of $\theta''$ precipitates that can be sheared by dislocations \citep{Bellon2020}. The DDD simulations took into account the main physical mechanisms that influence the dislocation/precipitate interactions (elastic mismatch between matrix and precipitate, transformation strains, cross-slip) and the simulation domain included a random and homogeneous distribution of precipitates following the experimental size distribution and oriented along the appropriate crystallographic directions. The dislocations were allowed to shear the precipitates if the driving force overcame a critical friction stress. The results of the simulations are in good agreement with the experimental results of the CRSS and  confirm that the maximum strength of the alloy is obtained when dislocation propagation throughout the microstructure involved a mixture of shearing of small precipitates and formation of Orowan loops around large precipitates.

\section{Material and experimental information}
\label{sec:Mat_and_exp}

\begin{figure}[t!]
\centering
\includegraphics[width=0.49\textwidth]{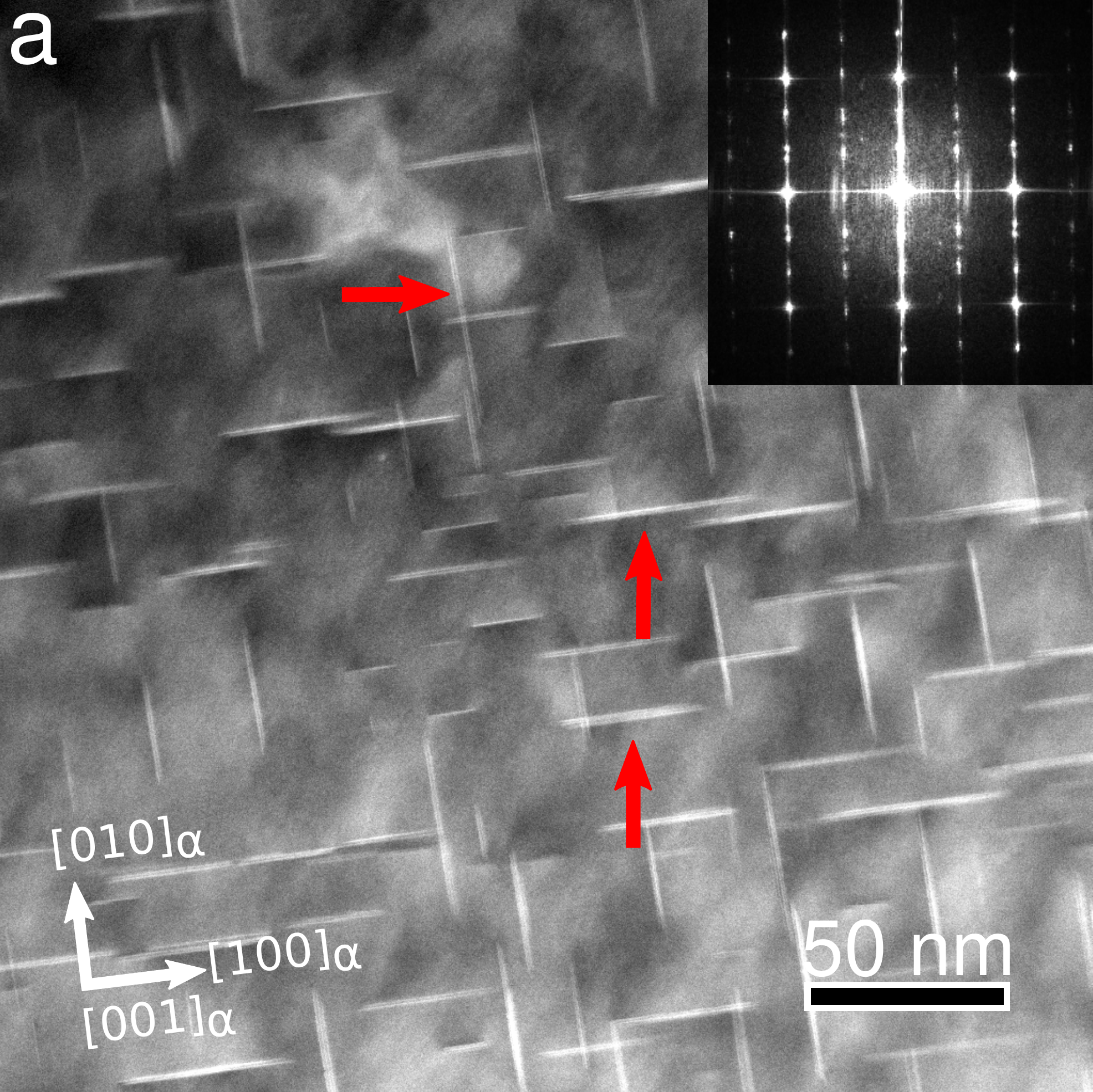} \\
\includegraphics[width=0.49\textwidth]{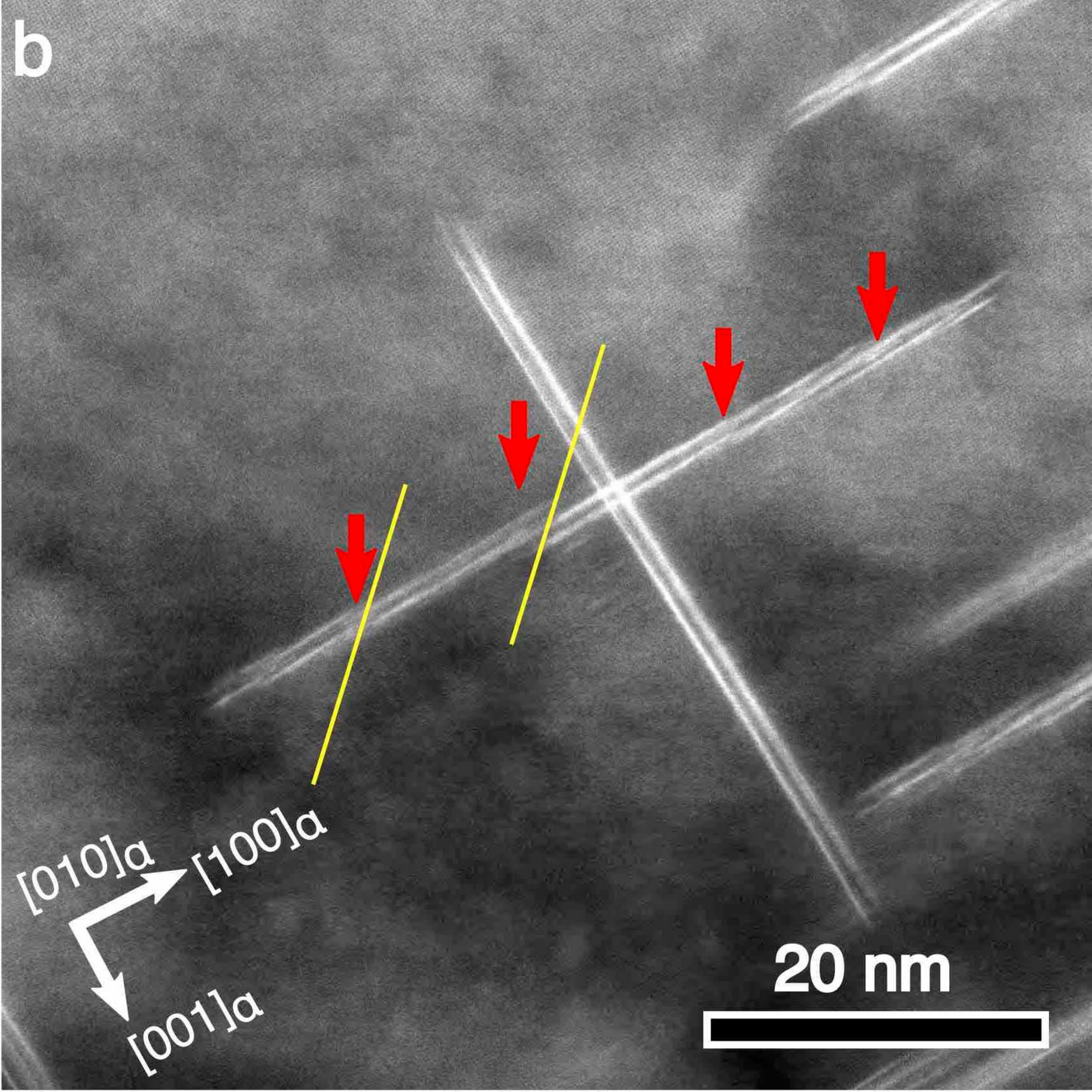}
\includegraphics[width=0.49\textwidth]{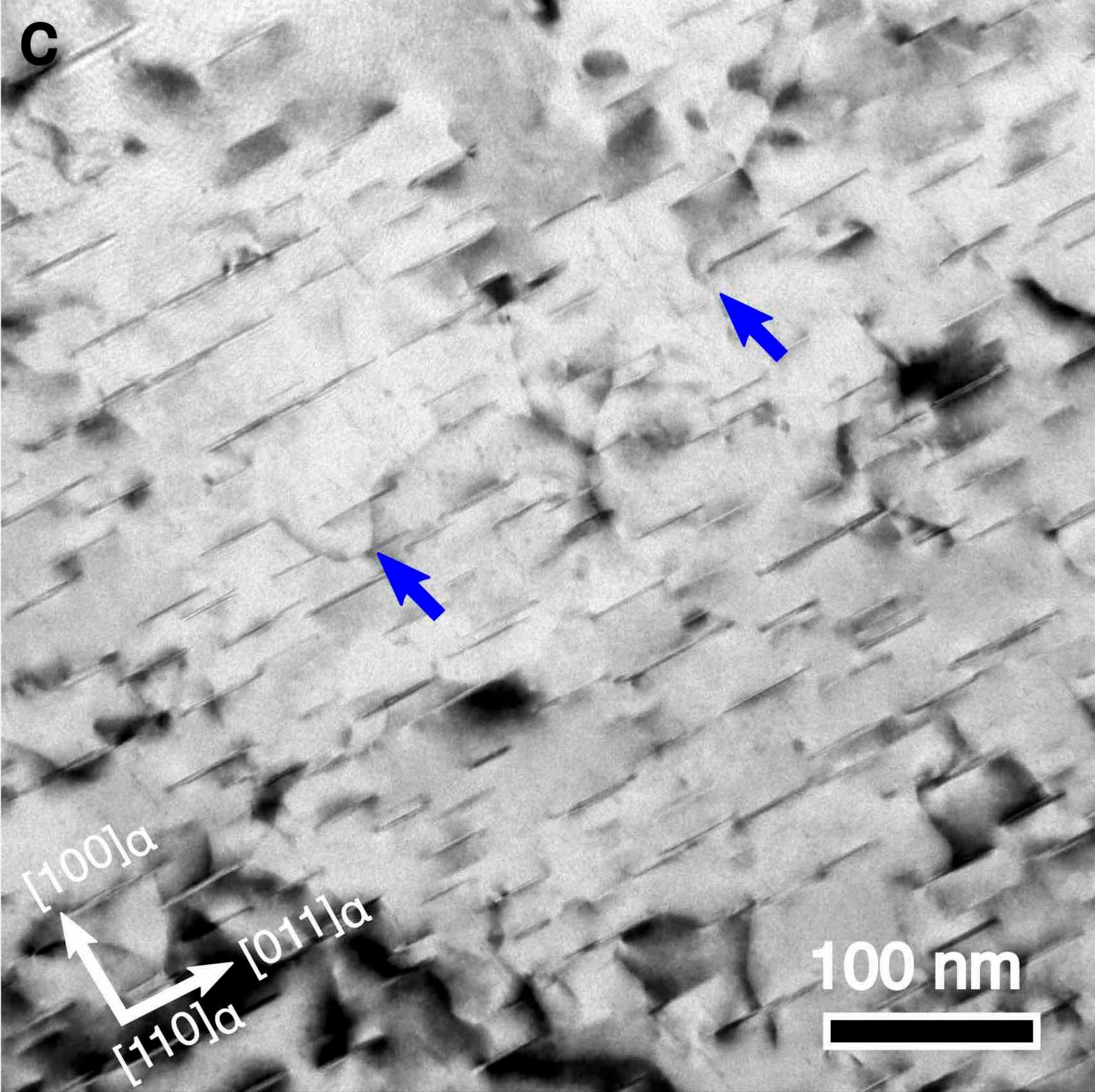}
\caption{(a) High-angle annular dark-field scanning-transmission electron micrograph showing the homogeneous distribution of $\theta''$ precipitates. The zone axis was [001]$_\alpha$. The Fast Fourier Transform (FFT) of the $\theta''$ precipitates is shown in the inset. (b) {\it Idem} showing $\theta''$ precipitates (diameter around 40 nm) sheared by dislocations. The red arrows indicate the planes in which the layers of Cu atoms have been disrupted by the shearing of the dislocation. The direction of the Burgers vector ([101]) is indicated by the yellow lines. (c) High-resolution transmission electron micrograph showing the formation of Orowan loops (blue arrows) between large $\theta''$ precipitates (diameter above 50 nm). Figures adapted from \cite{Bellon2020}.}  
\label{fig:TEM_30h}
\end{figure}

 The details of the processing, microstructure and properties of the Al - 4 wt. \% Cu alloy  can be found in \citep{Rodriguez-Veiga2018,Bellon2020} and only the most relevant information is included here for the sake of completion. The homogenized and solution treated samples were aged at $180^{\circ}$C for 30 hours leading to a homogeneous distribution of $\theta''$ precipitates. These precipitates are circular disks which grow parallel to the  {\{}001{\}}$_\alpha$ planes of the $\alpha$-Al matrix with an orientation relationship  (001)$_{\alpha}$//(001)$_{\theta''}$ and [001]${_\alpha}$//[001]$_{\theta''}$ (Fig. \ref{fig:TEM_30h}a). The diameter, thickness and volume fraction of the precipitates was carefully characterized from transmission electron microscopy micrographs. The volume fraction was 0.9 $\pm$ 0.2 \% while the thickness was 1.6 $\pm$ 0.2 nm. The precipitate diameter followed a log-normal distribution of mean $\mu$ = 3.389 and standard deviation $\sigma$ = 0.619, which correspond to an average diameter of  36 $\pm$ 26 nm.
 
Micropillars of  5 x 5 $\mu$m$^2$ square cross-section were manufactured by focus ion beam (FIB) milling. The aspect ratio of the micropillars (length/side) was in the range 2 to 3. An initial CRSS of  91 $\pm$ 5 MPa was measured from compression tests of the single crystal micropillars oriented for single slip. It was checked that the dimensions of the micropillars were large enough to eliminate size effects on the mechanical response. In addition, thin foils parallel to the  (001)$_\alpha$ and (100)$_\alpha$ were extracted from the deformed micropillars by FIB milling and examined in the transmission microscope to ascertain the dislocation/precipitate interactions. It was found that dislocations sheared small ($\le$ 40 nm) $\theta''$ precipitates (Fig. \ref{fig:TEM_30h}b) and formed Orowan loops around larger precipitates (Fig. \ref{fig:TEM_30h}c). More information about the microstructure and deformation mechanisms can be found in \cite{Bellon2020}.

\section{$\theta''$ precipitate features}
\label{sec:precipitate}

\begin{figure}
\centering
\includegraphics[width=0.49\textwidth]{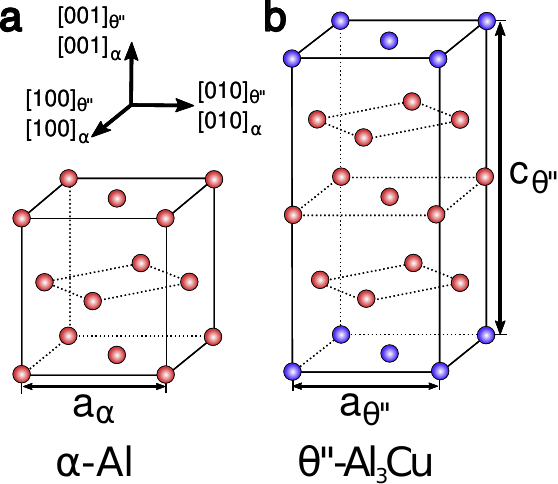}
\caption{Unit cells of $\alpha-$Al matrix and $\theta''$ precipitates \citep{Liu2019}. Cu atoms are blue and Al atoms red.}  
\label{fig:Unit_cell}
\end{figure}

$\theta''$ precipitates are a metastable phase that tends to nucleate homogeneously in the supersaturated $\alpha$-Al matrix during high temperature aging \citep{LML20, Liu2019}. The  $\theta''$ phase has a face center tetragonal lattice with a chemical composition Al$_3$Cu.  The corresponding unit cells of $\alpha$ Al and $\theta''$ are depicted in Fig. \ref{fig:Unit_cell}, which shows that $\theta''$ is formed by (001) layers of Cu atoms separated by three (001) layers of Al atoms. The lattice parameters of $\alpha$-Al  and $\theta''$ are $a_\alpha$ = $a_{\theta''}$ = 0.405 nm and $c_{\theta''}$ = 0.768 nm \citep{Liu2019}, leading to coherent interfaces along the (100) and (001) planes. Obviously, three different precipitate variants (associated with the cubic symmetry of the \{001\} planes) can be formed during precipitation. It should also be noted that dislocations gliding in the  \{111\}$<$110$>$ matrix slip system can penetrate easily into the precipitate without changing the Burgers vector due to the similarity of the lattices and the coherency of the interfaces.

The coherency strains in the matrix and in the precipitate due to the lattice mismatch can be determined from the transformation matrix, $\mathbf{T}$, that relates the lattice parameters in $\alpha$-Al, $\mathbf{e}_\alpha$, and in  $\theta''$,  $\mathbf{e}_{\theta''}$ ($\mathbf{T} \mathbf{e}_\alpha = \mathbf{e}_{\theta''}$). It can be computed as \citep{Liu2019}:

\begin{equation}
\begin{array}{ccc}
\mathbf{T}=\begin{pmatrix}
    a_{\theta''}/a_\alpha &0&0\\  0&a_{\theta''}/a_\alpha&0\\ 0&0&c_{\theta''}/2a_\alpha
  \end{pmatrix}
\end{array}
\label{eq:T}
\end{equation}

It can be observed that this transformation matrix $\mathbf{T}$ only contains dilatational components, as opposed to the case of $\theta '$ precipitates, in which there is an important shear component \citep{Liu2017}. 
The transformation matrix can be used to calculate the corresponding stress-free transformation strain (SFTS) of the $\theta ''$ precipitate, $\boldsymbol{\epsilon}^0$, according to:

\begin{equation}
\boldsymbol{\epsilon}^0 = \frac{1}{2}({\mathbf T}^T {\mathbf T} - {\mathbf I})
\label{eq:SFTS}
\end{equation}

\noindent where ${\mathbf I}$ stands for the identity matrix. Only one term of the transformation strain matrix was different from 0 and corresponded to a dilatational strain of  -5.05\%.  Therefore, the dilatational coherency strain is small and, thus, is expected to play a minor role in the precipitate strengthening.

The elastic constants of the $\alpha$-Al matrix and of the $\theta''$ precipitates were determined in previous investigations using first principles calculations and are shown in Tables \ref{Tab:ECA} and \ref{Tab:ECT} respectively \citep{Rodriguez-Veiga2018}.

\begin{center}
\begin{table}[h]
\begin{center}
\begin{tabular}{cccc}
\hline
  $C_{11}$ & $C_{12}$ & $C_{44}$\\
\hline
 110.4  & 60.0 & 31.6 \\
\hline
\end{tabular}
\end{center}
\caption{Elastic constants (in GPa) of $\alpha$ - Al obtained from first principles calculations.}
\label{Tab:ECA}
\end{table}
\end{center}

\begin{center}
\begin{table}[h]
\begin{center}
\begin{tabular}{cccccc}
\hline
 $C_{11}=C_{22}$ & $C_{12}$ & $C_{13}=C_{23}$ & $C_{33}$ & $C_{44}=C_{55}$ & $C_{66}$\\
\hline
 160.8 & 56.5 & 48.4 & 175.6 & 46.2 & 50.1\\
\hline
\end{tabular}
\end{center}
\caption{Elastic constants (in GPa) of  $\theta''$ - Al$_{2}$Cu precipitates obtained from first principle calculations. }
\label{Tab:ECT}
\end{table}
\end{center}

\section{Discrete dislocation dynamics framework}
\label{sec:Methodology}

The interaction of dislocations with precipitates was simulated using an updated version of the DDD framework employed in \cite{Santos-Guemes2018} and \cite{Santos-Guemes2020}. The dislocation loops are discretized in segments and the displacement of each segment in each simulation step depends on the Peach-Koehler force. The plastic strain is computed directly from the area sheared by the dislocation loop and this information is used to determine the mechanical fields in the simulation domain. The mechanical fields in each simulation step are efficiently obtained in each simulation step through a Fast Fourier Transform (FFT) solver, leading to an efficient DDD tool for problems involving elastic and plastic heterogeneities as well as eigenstrains \citep{Bertin2018}. More details about the DDD tool can be found in our previous publications \citep{Santos-Guemes2018,Santos-Guemes2020,Bertin2018} and only the novel features added to the tool -regarding dislocation mobility rules in the matrix and the precipitate,  computation of the mechanical fields and shearing of the precipitates- are detailed below.

\subsection{Dislocation mobility}

The dislocation lines are discretized in segments limited by nodes. The displacement of each node depends on the segment mobility and the Peach-Koehler force, which in turn depends on the local stress state at the dislocation segments. Once the full mechanical fields are computed following the procedure indicated in the next section, the nodal force $\mathbf{F}_i$ is determined as

 \begin{equation}
\mathbf{F}_i = \sum_j \mathbf{f}_{ij}
\label{Fi}
\end{equation}

\noindent where $\mathbf{f}_{ij}$ is the force acting on the segment $ij$ (limited by nodes $i$ and $j$), which is computed according to

 \begin{equation}
\mathbf{f}_{ij} = \int_{\mathbf{x}_i}^{\mathbf{x}_j} N_i(\mathbf{x}) \mathbf{f}_{ij}^{pk}(\mathbf{x}) \rd \mathbf{x}
\label{fij}
\end{equation}

\noindent where $N_i$ is the interpolation function associated with node $i$ and $\mathbf{f}_{ij}^{pk}$ is the Peach-Koehler force given by

 \begin{equation}
\mathbf{f}_{ij}^{pk}(\mathbf{x}) =  \Big(\boldsymbol{\sigma}(\mathbf{x}) \cdot  \mathbf{b}_{ij}\Big) \times \mathbf{\hat t}_{ij} 
\label{fijpk}
\end{equation}

\noindent where $\mathbf{b}_{ij}$ is the Burgers vector of the segment $ij$ and $\mathbf{\hat t}_{ij}$ the unit vector parallel to the dislocation line.

The velocity of node $i$ in the glide plane of the matrix, $\mathbf{v}_i^M$,  is given by

\begin{equation} \label{eq:mobility}
\mathbf{v}_i^M = \left\{
 \begin{array}{ll}
 [\mathbf{F}_i^{g}-F_i^{M}(\mathbf{F}_i^{g}/|\mathbf{F}_i^{g}|)]/B^M & \mathrm{if} \quad |\mathbf{F}_i^{g}| > F_i^{M}
\\ \\
 0 &  \mathrm{if} \quad|\mathbf{F}_i^{g}| \leq F_i^{M}
 \end{array}
 \right.
\end{equation}

\noindent while the velocity of node $i$ in the glide plane of the precipitate, $\mathbf{v}_i^P$,  is given by

\begin{equation} \label{eq:mobilityP}
\mathbf{v}_i^P = \left\{
 \begin{array}{ll}
 [\mathbf{F}_i^{g}-F_i^{P}(\mathbf{F}_i^{g}/|\mathbf{F}_i^{g}|)]/B^P & \mathrm{if} \quad |\mathbf{F}_i^{g}| > F_i^{P}
\\ \\
 0 &  \mathrm{if} \quad|\mathbf{F}_i^{g}| \leq F_i^{P}
 \end{array}
 \right.
\end{equation}

\noindent where $\mathbf{F}_i^{g}$ is the projection of the nodal force, $\mathbf{F}_i$, on the glide plane (characterized by the slip plane normal $\mathbf{n}$) according to

\begin{equation}
\mathbf{F}_{i}^{g} =  \mathbf{F}_{i} - (\mathbf{F}_{i} \cdot \mathbf{n})\mathbf{n}.
\end{equation}

\noindent $F_i^{M}$  and  $F_i^{P}$ stand for force thresholds due to friction in the matrix and to resistance to shearing of the precipitate, respectively.  In the case of FCC $\alpha$-Al, the Peierls stress is negligible and only solid solution contributes to the matrix friction. These force thresholds can be determined from eqs. \eqref{Fi}  to  \eqref{fijpk}  by assuming that the Peach-Koehler force in each dislocation segment accounting for either matrix friction, $f^{pk,M}_{ij}$, or the precipitate resistance to shearing,  $f^{pk,P}_{ij}$, are given by

\begin{equation}
f^{pk,M}_{ij}=\tau^{ss}b \qquad {\rm and} \qquad  f^{pk,P}_{ij}=\tau^{P}b
\end{equation}

\noindent where $\tau^{ss}$ is contribution of solid solution to the CRSS and $\tau^P$ stands for the threshold shear stress necessary to shear the precipitates and the same Burgers vector $b$ was assumed for the matrix and the precipitate. These approaches to take into account solid solution strengthening and the shear resistance of the precipitates have been used by  \cite{Queyreau2010_superposition} and \cite{monnet2015multiscale}, respectively, within the framework of DDD simulations.

Finally, $B^M$  and $B^P$ are viscous drag coefficients that control the mobility of the dislocation in the matrix and in the precipitate, respectively, and that depend on the dislocation character. $B^M$  was obtained for Al from atomistic simulations \citep{Molinari2017_Mobility,Santos-Guemes2018} but $B^P$ is not known and it was assumed that  $B^P$ = $B^M$. This assumption was made after checking that precipitate shearing was mainly controlled by the friction stress $\tau^P$ and large changes in $B^P$ did not influence the CRSS of the alloy. Equations of motion were integrated using an Euler explicit scheme.

Dislocation cross-slip was included through a probabilistic, thermally-activated model that takes into account the local stress state on the dislocation segments 
\citep{Esteban-Manzanares2020_CS,Santos-Guemes2020}. The cross-slip probability $P$ of a screw dislocation of length $L$ is computed according to 

\begin{equation}
P(\tensor{\sigma},T) = \nu_{eff}\frac{L}{L_n}e^{-\left[\Delta H(\tensor{\sigma})\left(1-\frac{T}{T_m}\right)\right]/k_bT}\Delta t
\label{eq:CS_prob}
\end{equation}

\noindent where $\nu_{eff}$ is the attempt frequency, $k_b$ stands for the Boltzmann constant, $\Delta t$ is the time step, $L_n$ is the nucleation length and $T$ and $T_m$ are the temperature of the simulation and the melting temperature of the material respectively. Finally, $\Delta H$ is the activation enthalpy for cross-slip as function of the local stress state. These parameters were obtained for Al from atomistic simulations \cite{Esteban-Manzanares2020_CS} and the details of the cross-slip implementation as well as the values of the different parameters in eq. \eqref{eq:CS_prob} can be found in \cite{Santos-Guemes2020}.

\subsection{Computation of mechanical fields} 
 
The stress field within the simulation domain is computed using the FFT algorithm by solving the following mechanical equilibrium equations with periodic boundary conditions in the simulation domain $V$
 
\begin{equation} \label{eq:MechEq}
 \left\{
 \begin{array}{l}
  \tensor{\sigma}(\mathbf{x})=\ftensor{C}(\mathbf{x}):[\tensor{\epsilon}(\mathbf{x})-\tensor{\epsilon}^p(\mathbf{x})-\delta(\mathbf{x})\tensor{\epsilon}^0], \quad \forall \mathbf{x} \in V
\\
 \mathrm{div} ( \tensor{\sigma}(\mathbf{x}))=0  \quad \mathbf{x} \in V \\
\tensor{\sigma}  \cdot \mathbf{n} \text{ opposite on opposite sides of } \partial V \\
\frac{1}{V}\int_V \tensor{\epsilon}(\mathbf{x})=E
 \end{array}
 \right.
\end{equation}\\

\noindent where $\ftensor{C}$ denotes the fourth order elasticity tensor, $\tensor{\epsilon}$ the total strain, $\tensor{\epsilon}^p$ the plastic strain, $\tensor{\epsilon}^0$ the stress-free transformation strain associated to the precipitate and $\partial V$ stands for the boundaries of domain $V$ with normal $\mathbf{n}$ and $E$ is the imposed macroscopic strain. $\delta(\mathbf{x})$ is a Dirac delta function that is equal to 1 when $\mathbf{x}$ is within the precipitate and 0 otherwise.

In the previous versions of the DDD tool \citep{Capolungo2015_DDDFFT, Bertin2018,Santos-Guemes2018,Santos-Guemes2020}, the plastic strain $\tensor{\epsilon}^p$ was computed directly from the motion of dislocations, following the eigenstrain-based formalism of the discrete continuum approach developed by \cite{Lemarchand2001_DCM}. Recently, \cite{Bertin2019} introduced a new strategy to compute  $\tensor{\epsilon}^p$ based on a spectral-based implementation of the Field Dislocation Mechanics (FDM) approach \citep{Djaka17}. The foundation of the FDM is the dislocation density tensor $\tensor{\alpha}$ introduced by \cite{Nye53}, $\alpha_{ij}=b_it_j$, where $b_i$ is the net Burgers vector in direction $\mathbf{e}_i$ per unit surface and $t_j$ the dislocation line direction along $\mathbf{e}_j$. From the dislocation density tensor $\tensor{\alpha}$, the FDM approach yields the following Poisson-type equation:

\begin{equation}
\rm{div}(\rm{grad}(\mathbf{U}^{e,\bot}))=-\rm{curl} (\tensor{\alpha})
\label{eq:Poisson_FDM}
\end{equation}

\noindent where $\mathbf{U}^{e,\bot}$ stands for the incompatible part of the elastic distortion. Eq. \eqref{eq:Poisson_FDM} can be efficiently solved in the Fourier space obtaining the field $\mathbf{U}^{e,\bot}$, whose symmetric part is the tensor $\tensor{\epsilon}^p$ that can be introduced in the mechanical equilibrium  equation, eq. \eqref{eq:MechEq}. The value of the transformation strain was smoothened and spread to nearby voxels  in contact with the precipitates to reduce fluctuations in the computed mechanical fields that may appear due to the FFT solver. 
More detailed information about the FDM and FFT solver can be found in \citep{Djaka17,Santos-Guemes2018,Bertin2018,Bertin2019,KC20}.

\subsection{Matrix/precipitate interfaces}

\begin{figure}
\centering
\includegraphics[width=0.7\textwidth]{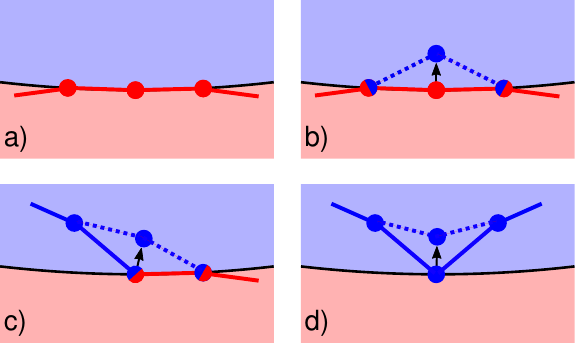}
\caption{Schematic of the dislocation penetration through the interface between two phases, one represented in red and another  in blue. Solid and dashed lines show the dislocation segments before and after  node transfer across the interface, respectively.  (a) Impenetrable precipitate. (b) Node transfer with three consecutive nodes on the interface. (c) Node transfer with two consecutive nodes on the interface. (d) Node transfer with only one node on the interface.}  
\label{fig:Transfer_node}
\end{figure}

Introduction of both shearable and non-shearable second phases within the simulation domain requires a careful treatment of the matrix/precipitate interfaces, which should allow to move (or not)  dislocation nodes through the interface. In a general case,  the dislocation line is initially gliding in a slip system of one phase until it reaches the interface with the second phase (Fig. \ref{fig:Transfer_node}a). At this point, it has to be evaluated whether the dislocation can penetrate into the precipitate according to eq. \eqref{eq:mobilityP}. If this condition is not fulfilled, the precipitate is treated as impenetrable and the dislocation nodes are forced to remain on the interface (Fig. \ref{fig:Transfer_node}a). Otherwise, node transfer through the interface and different scenarios have to be considered. If three consecutive nodes are on the interface (Fig. \ref{fig:Transfer_node}b), the middle node and the two adjacent segments are transferred to the slip system in the precipitate which is equivalent to the slip system in the matrix. The other end nodes of the segments are mapped to the interface. If one of the adjacent segments of the node is already in the second phase (Fig. \ref{fig:Transfer_node}c), the other node can be transferred to the second phase provided that the third node is on the interface. Finally, if both connections of one node on the interface are in one phase, this node can be transferred to that phase (Fig. \ref{fig:Transfer_node}d).

Voxels belonging to the precipitate are assigned the precipitate properties, that include the extra friction stress $\tau^P$ that models the precipitate resistance to be sheared by the dislocation. 

\section{Results and experimental validation}
\label{sec:Results}

\subsection{Simulation details}

\begin{figure}[t!]
\centering
\includegraphics[width=0.49\textwidth]{Edge3D_green.pdf}
\includegraphics[width=0.49\textwidth]{Screw3D_green.pdf}
\includegraphics[width=0.49\textwidth]{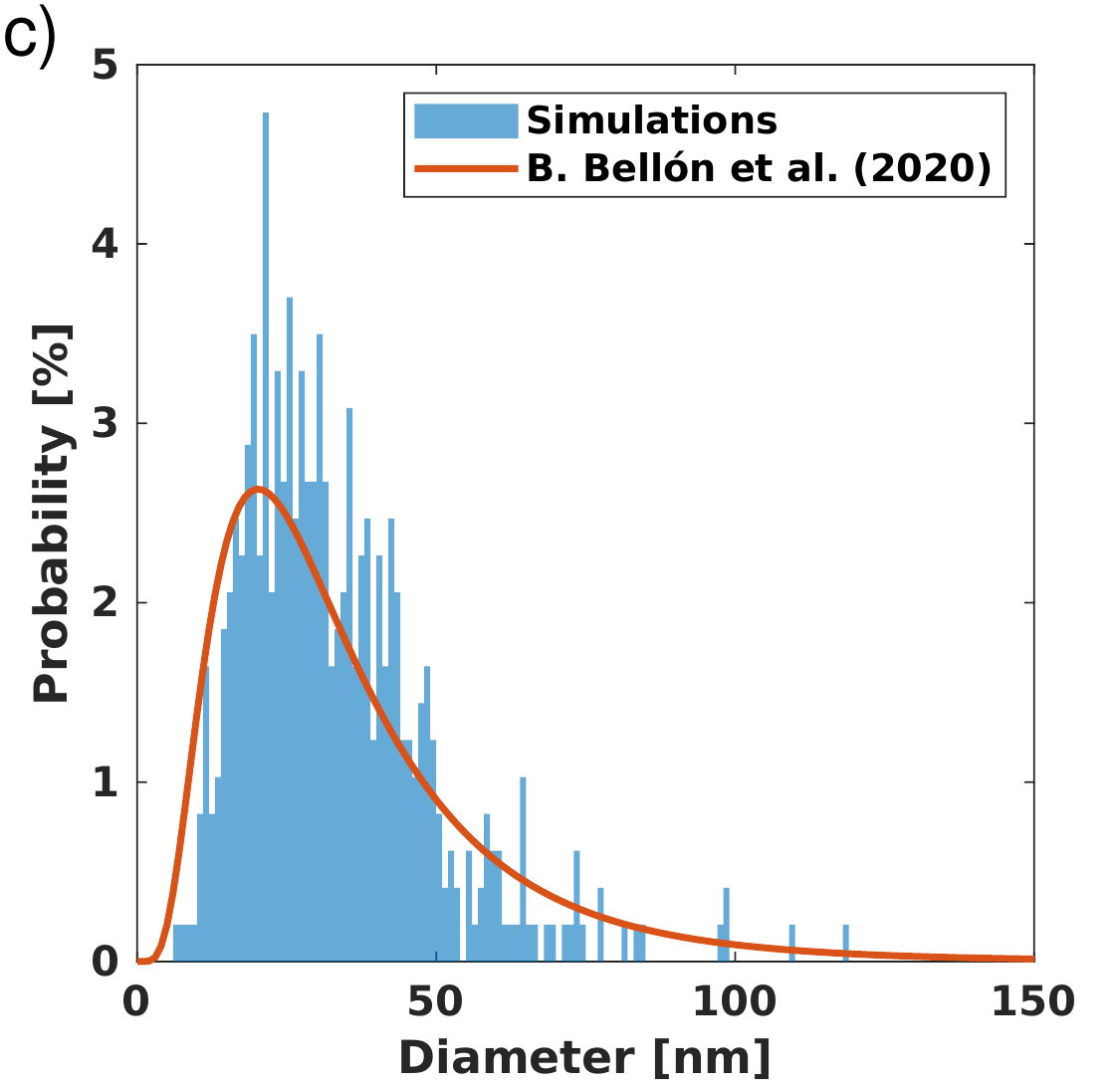}
\caption{Simulation domains containing a lognormal distribution of $\theta''$ precipitates. (a) Initial edge dislocation. (b) Initial screw dislocation. The color of the precipitates indicates the orientation variant. Initial dislocations are represented by a straight black line. (c) Size distribution of the precipitates in the simulation domains. The red curve corresponds to the experimental one \citep{Bellon2020}.}  
\label{fig:Distributions}
\end{figure}

DDD simulations were carried out in a cubic  domain of 200 x 200 x 200 nm$^3$ with periodic boundary conditions. The domain was discretized with a grid of 128 x 128 x 128 voxels. Both initial edge and screw dislocations were considered. For the case of an initial edge dislocation, the axes of the cubic domain were aligned with the [1$\bar1$2], [110] and [$\bar1$11] directions of the $\alpha$-Al lattice, and an straight edge dislocation ($\bar1$11)[110] was introduced in the domain (Fig. \ref{fig:Distributions}a). The axes of the cubic domain in the case of the initial screw dislocation were aligned with the [110], [$\bar1$1$\bar2$] and [$\bar1$11] directions of the $\alpha-$Al lattice and a straight screw dislocation ($\bar1$11)[110] was introduced (Fig. \ref{fig:Distributions}b). These configuration led to an initial dislocation density of $2.5\cdot10^{13}$ m$^{-2}$, very close to that measured in the micropillars before deformation ($2.2\cdot10^{13}$ m$^{-2}$) \citep{Bellon2020}. 

Ten different precipitate distributions were randomly generated within the simulation domain. The precipitates have the shape of circular disks with the broad face of the precipitate parallel to either the (100), (010) or (001) planes of the FCC $\alpha$-Al lattice to account for the three different precipitate variants. The thickness of the precipitates was constant and equal to 1.6 nm, while the diameters were randomly selected following the experimental lognormal distribution \citep{Bellon2020}. The actual precipitate size distribution used in the simulations is compared with the experimental one in Fig. \ref{fig:Distributions}c. The volume fraction of precipitates was 1{\%}. Approximately 30 to 70 precipitates were included in each cubic domain.  An edge dislocation was introduced in five of the precipitate realizations and a screw in another five. Two examples of the distributions are depicted in Fig. \ref{fig:Distributions}.  The experimental evidence indicates that $\theta''$ precipitates can be sheared by dislocations \citep{Bellon2020} but the threshold stress necessary to shear the precipitates is unknown. Therefore, simulations with three different threshold stress for precipitate shearing, $\tau^P$ = 1 GPa, 2 GPa and $\infty$, were performed to ascertain the influence of this parameter on the CRSS. It should be noted that previous attempts to estimate the threshold stress of nm-sized Cu and Cr precipitates in Fe using atomistic simulations \citep{TBM08, monnet2015multiscale} led  to values in the range 2 to 2.5 GPa. Finally,  the effect of solid solution was included in the analysis through the friction stress $\tau^{ss}$ which depends on volume fraction of Cu
atoms which remain in solid solution after the heat treatment. Taking into account the precipitate
volume fraction of 1\%, the remaining fraction of Cu atoms in solid solution was 3.5\%  and $\tau^{ss}$ = 25.7 MPa according to \cite{Bellon2020}.

 The simulation features (domain size, initial dislocation density, precipitate size and content, strain rate, É) were set to achieve conditions that are representative of those found in the micropillar compression experiments, in which the CRSS was independent of the micropillar dimensions and the micropillars deformed by single slip along parallel slip planes. Thus, a shear strain parallel to the $(\bar{1}11)$ plane in the [110] direction red was applied to generate the driving force required to move the dislocations in the $(\bar{1}11)$ glide plane.

Micropillar compression experiments in smaller micropillars (2 x 2 $\mu$m$^2$) showed as size effect of the type "smaller is stronger" \citep{Bellon2020}. This size effect was attributed the limited density of mobile dislocations within the micropillar, that were not able to accommodate the strain applied to the micropillar. As a result, part of the strain has to be accommodated by elastic strains, leading to elastic hardening. In order to avoid this size effect in the simulations, the applied strain rate has to be sufficiently small to allow the accommodation of the applied strain by dislocation slip. The effect of the strain rate on the CRSS was analyzed in the simulations, and it was found to be negligible for strain rates below or equal to 10$^5$ s$^{-1}$ for the domain size and precipitate volume fraction considered in this investigation. Therefore, simulations were performed at a strain rate of 10$^5$ s$^{-1}$  to optimize the computational efficiency.

Finally, the dislocation/dislocation interactions in the experiments were mainly due to their stress fields  and the interaction of the moving dislocations with forest dislocations was limited because of the low initial dislocation density. The elastic interactions between the stress fields of dislocations gliding in parallel slip planes were taken into account in the simulations since infinite replicas of the box are considered through the periodic boundary conditions. On the contrary, the interactions with forest dislocations were not accounted for. Nevertheless, the contribution  to the CRSS of mobile dislocation interactions with forest dislocations can be neglected in this case in comparison with the influence of the dislocation/precipitate interactions. 

\subsection{Dislocation/precipitates interactions}

\begin{figure}
\centering
\includegraphics[width=0.49\textwidth]{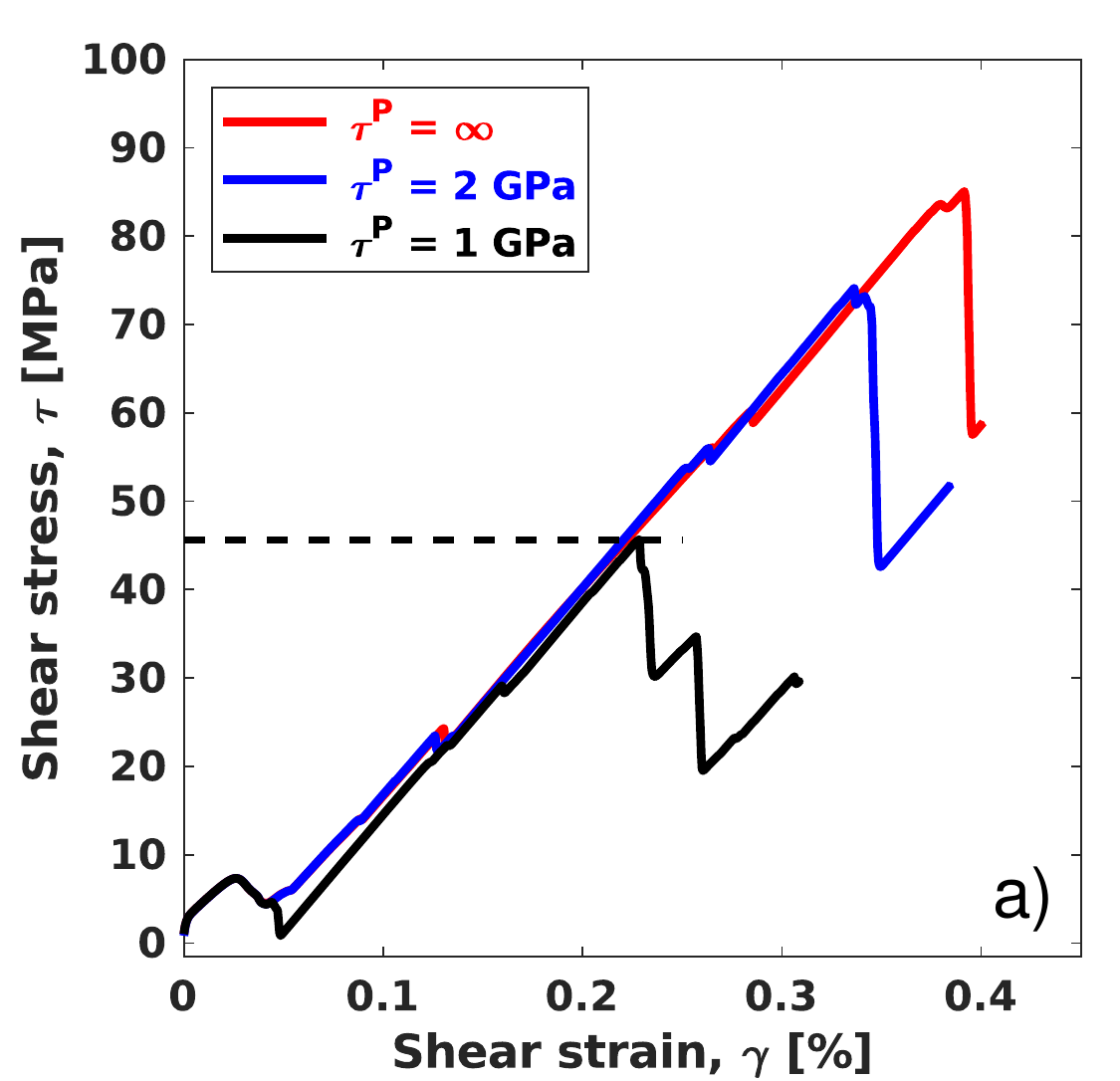}
\includegraphics[width=0.49\textwidth]{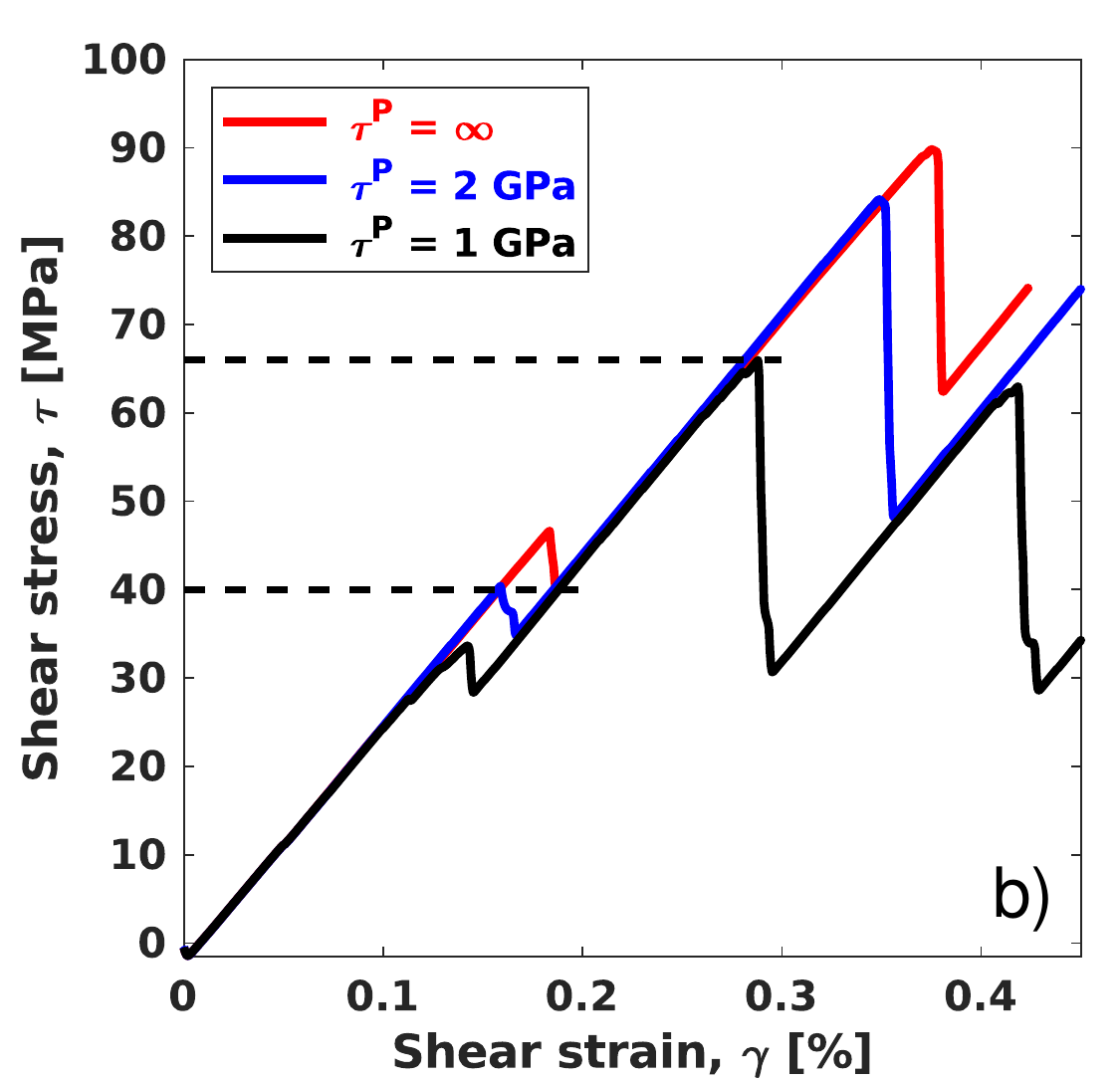}
\caption{Representative shear stress-shear strain curves obtained from DDD simulations with (a) initial edge dislocation and (b) initial screw dislocation. Results for shearable  ($\tau^P=1$ GPa and $\tau^P=2$ GPa) and impenetrable ($\tau^P=\infty$) precipitates are shown. The stress levels indicated with dashed black lines correspond to the snapshots depicted in Figs. \ref{fig:Snapshot_Edge04}, \ref{fig:Snapshot_Screw02_1} and \ref{fig:Snapshot_Screw02_2}}  
\label{fig:stress_strain_curve01}
\end{figure}

Simulations of all the precipitate distributions were performed  assuming that  the precipitates were impenetrable or could be sheared with $\tau^P=1$ GPa and $\tau^P=2$ GPa. Representative shear stress - shear strain curves of the propagation of an edge and a screw dislocation through a precipitate distribution are plotted in Figs. \ref{fig:stress_strain_curve01}a and b, respectively for the three sets of simulations. It should be noted that these simulations included the most relevant mechanisms that control the dislocation/precipitate interactions: elastic heterogeneity, coherency stresses due to the transformation strains and cross-slip. 

\begin{figure}[t!]
\centering
\includegraphics[width=\textwidth]{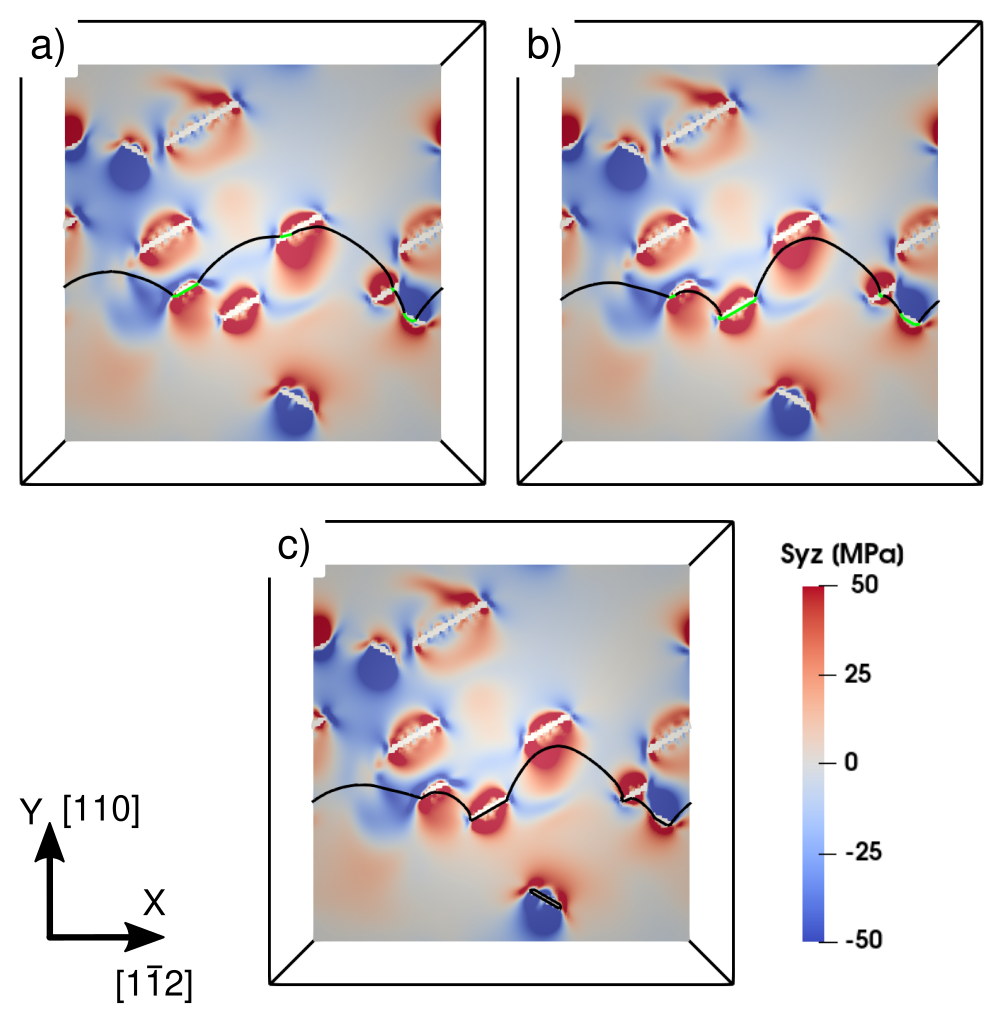}
\caption{Snapshot of the edge dislocation/precipitates interaction at  $\tau$ = 46 MPa of the stress-strain curve in Fig. \ref{fig:stress_strain_curve01}a. a) $\tau^P$ = 1 GPa. b) $\tau^P$ = 2 GPa.  c) $\tau^P$ = $\infty$. The shear stress field shown corresponds to the one created by the coherency strains. The dislocation lines in the matrix appear black and in the precipitate green.}  
\label{fig:Snapshot_Edge04}
\end{figure}

\begin{figure}[t!]
\centering
\includegraphics[width=\textwidth]{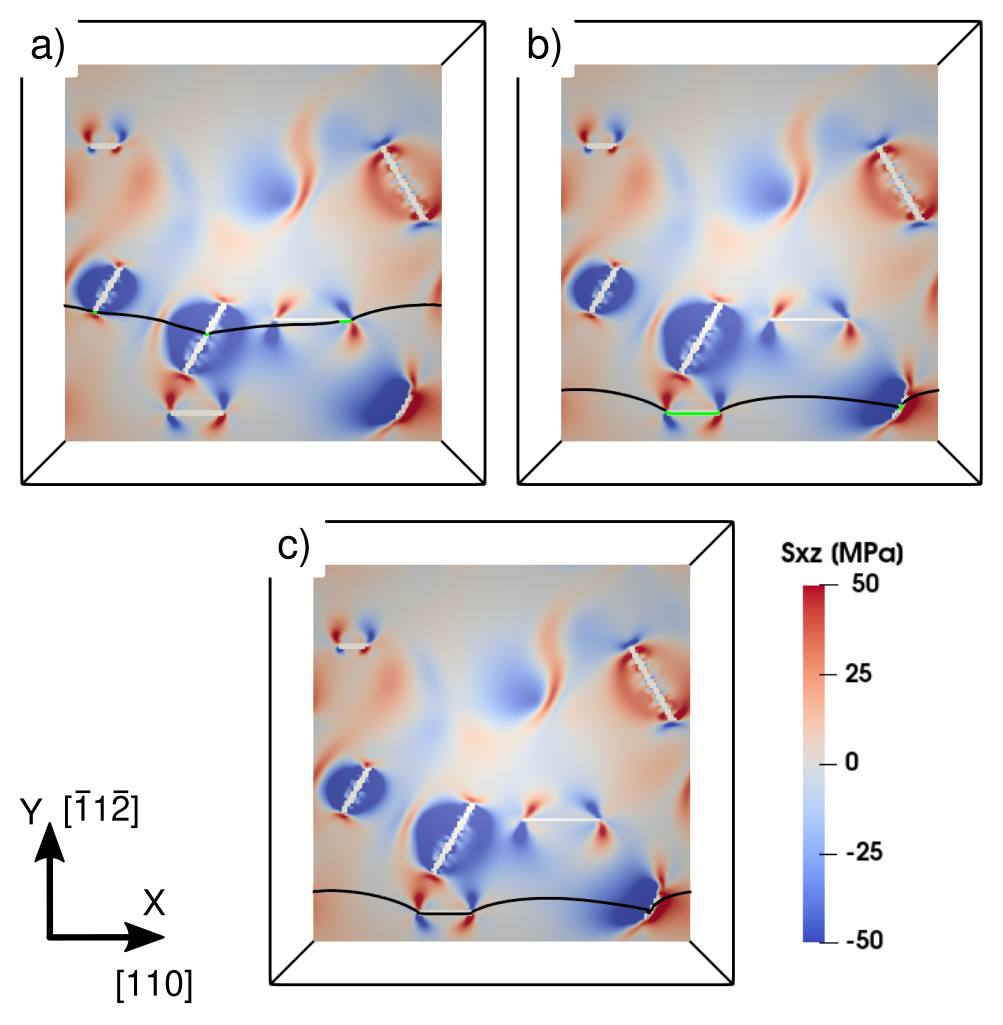}
\caption{Snapshot of the screw dislocation/precipitates interaction at stress $\tau$ =40 MPa of the stress-strain curve in Fig. \ref{fig:stress_strain_curve01}b. a) $\tau^P$ = 1 GPa. b) $\tau^P$ = 2 GPa.  c) $\tau^P$ = $\infty$. The shear stress field shown corresponds to the one created by the coherency strains. The dislocation lines in the matrix appear black and in the precipitate green.}   
\label{fig:Snapshot_Screw02_1}
\end{figure}

\begin{figure}[t!]
\centering
\includegraphics[width=\textwidth]{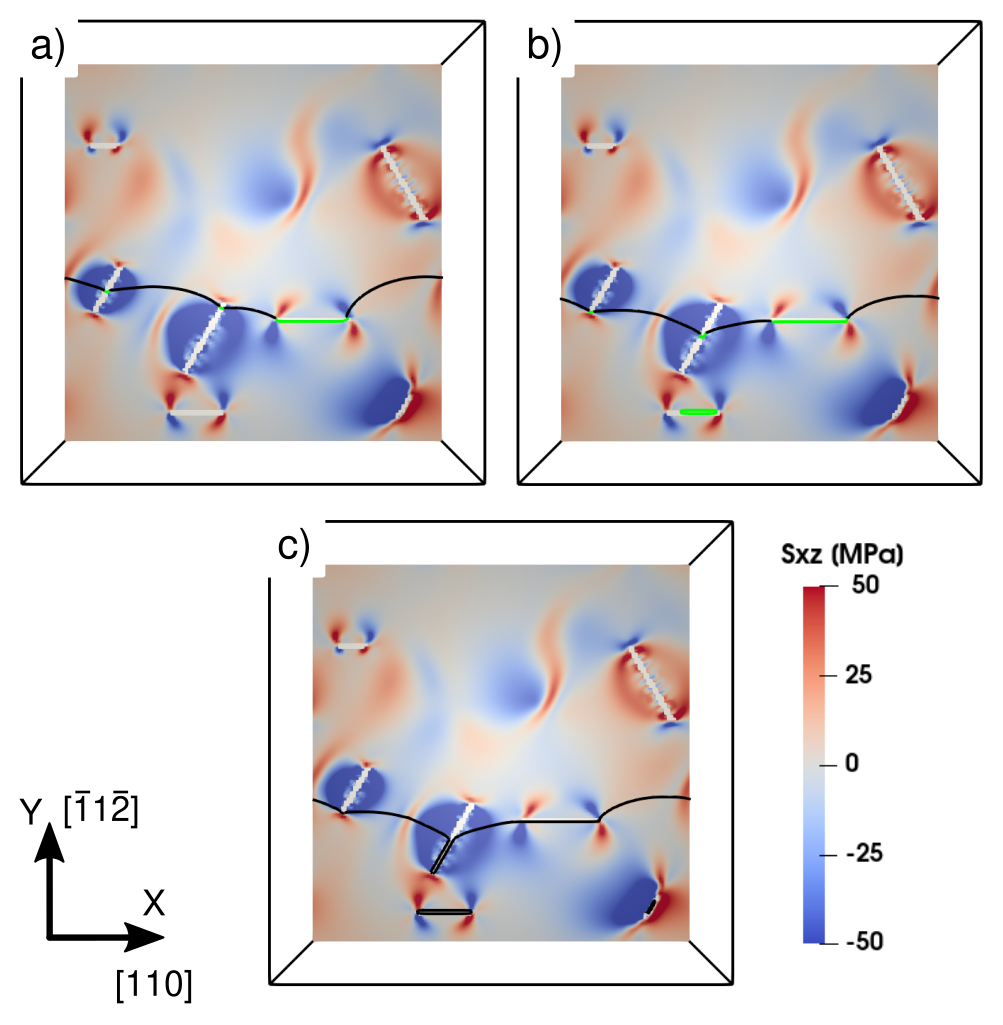}
\caption{Snapshot of the screw dislocation/precipitates interaction at stress $\tau$ =66 MPa of the stress-strain curve in Fig. \ref{fig:stress_strain_curve01}b. a) $\tau^P$ = 1 GPa. b) $\tau^P$ = 2 GPa.  c) $\tau^P$ = $\infty$. The shear stress field shown corresponds to the one created by the coherency strains. The dislocation lines in the matrix appear black and in the precipitate green.}    
\label{fig:Snapshot_Screw02_2}
\end{figure}

The progression of the dislocation line in the glide plane through the forest of precipitates is shown in the movies Edge.mp4 and Screw.mp4 in the Supplementary material for the curves depicted in Figs. \ref{fig:stress_strain_curve01}a and b, respectively. In addition, snapshots of the dislocation/precipitates interaction in the glide plane, corresponding to the stress levels indicated with dashed lines in Fig.\ref{fig:stress_strain_curve01}, are shown in Figs. \ref{fig:Snapshot_Edge04}, \ref{fig:Snapshot_Screw02_1} and \ref{fig:Snapshot_Screw02_2}. The shear stress field in the glide plane due to the coherency strains is shown in the movies and in the snapshots. The dislocation segments gliding in the matrix are depicted with black lines, whereas dislocation segments shearing the precipitates are plotted as green lines. The initial region of the shear stress-strain curve is linear (the small bump at the beginning of the curve in Fig. \ref{fig:stress_strain_curve01}a is due to the accommodation of the dislocation line to the coherency stresses around the precipitates) and the curves corresponding to shearable and impenetrable precipitates are practically superposed. However, differences in the curves appear when shear stress reaches $\approx$ 40 MPa. 

In the case of an initial edge dislocation, the dislocation line has sheared most of the precipitates in the central region of the glide plane when $\tau^P$ = 1 GPa (Fig. \ref{fig:Snapshot_Edge04}a) and the CRSS to overcome the precipitates in the glide plane has been attained because shearing of other precipitates ahead of the dislocation line will be carried out lower stresses. On the contrary, the dislocation was not able to overcome the precipitates in the central region of the glide plane by either shearing (Fig. \ref{fig:Snapshot_Edge04}b) or forming  an Orowan loop around the precipitates (Fig. \ref{fig:Snapshot_Edge04}c) when $\tau^P$ = 2 GPa or $\infty$, respectively, at this stress level. This was only accomplished after further straining and the CRSS to overcome the precipitates increased with   $\tau^P$. Similar behavior was observed in the case of a screw dislocation. The first peak in the shear stress - strain curves around $\approx$ 40 MPa (Fig. \ref{fig:stress_strain_curve01}b) is associated with a first row of two precipitates (Fig. \ref{fig:Snapshot_Screw02_1}) and the stress necessary to overcome these obstacles increases with $\tau^P$. The CRSS is attained when the dislocation overcomes the second row of precipitates in the center of the domain (Fig. \ref{fig:Snapshot_Screw02_2}) and is much smaller in the case of weak precipitates ($\tau^P$ = 1 GPa) that are easily sheared by the dislocation. It is worth noting at this point that an Orowan loop was formed around the precipitate parallel and close to the initial dislocation line in the case of impenetrable precipitates (Fig. \ref{fig:Snapshot_Screw02_2}c). A similar Orowan loop has been formed within this precipitate in the simulations with $\tau^P$ = 2 GPa (Fig. \ref{fig:Snapshot_Screw02_2}b), and this behavior indicates that $\tau^P$ = 2 GPa was close to the threshold stress to avoid precipitate shearing.

Simulations were carried out turning off each of the physical mechanisms that influence dislocation/precipitate interactions. It was found that the effect of the elastic mismatch between matrix and precipitates, the coherency stresses and  cross-slip on the dislocation path on the shear stress-strain curves was negligible. Previous investigations of the interaction of dislocations with $\theta'$ precipitates \citep{Santos-Guemes2020}, whose formation is associated with large transformation shear strains, showed that these transformation strains modified significantly the dislocation path through the precipitate forest and promoted cross-slip. Nevertheless, the out-of-plane driving forces due to the coherency strains and to the elastic mismatch were very small in the case of $\theta''$ precipitates and  were not able to promote dislocation cross-slip.

\begin{figure}
\centering
\includegraphics[width=0.49\textwidth]{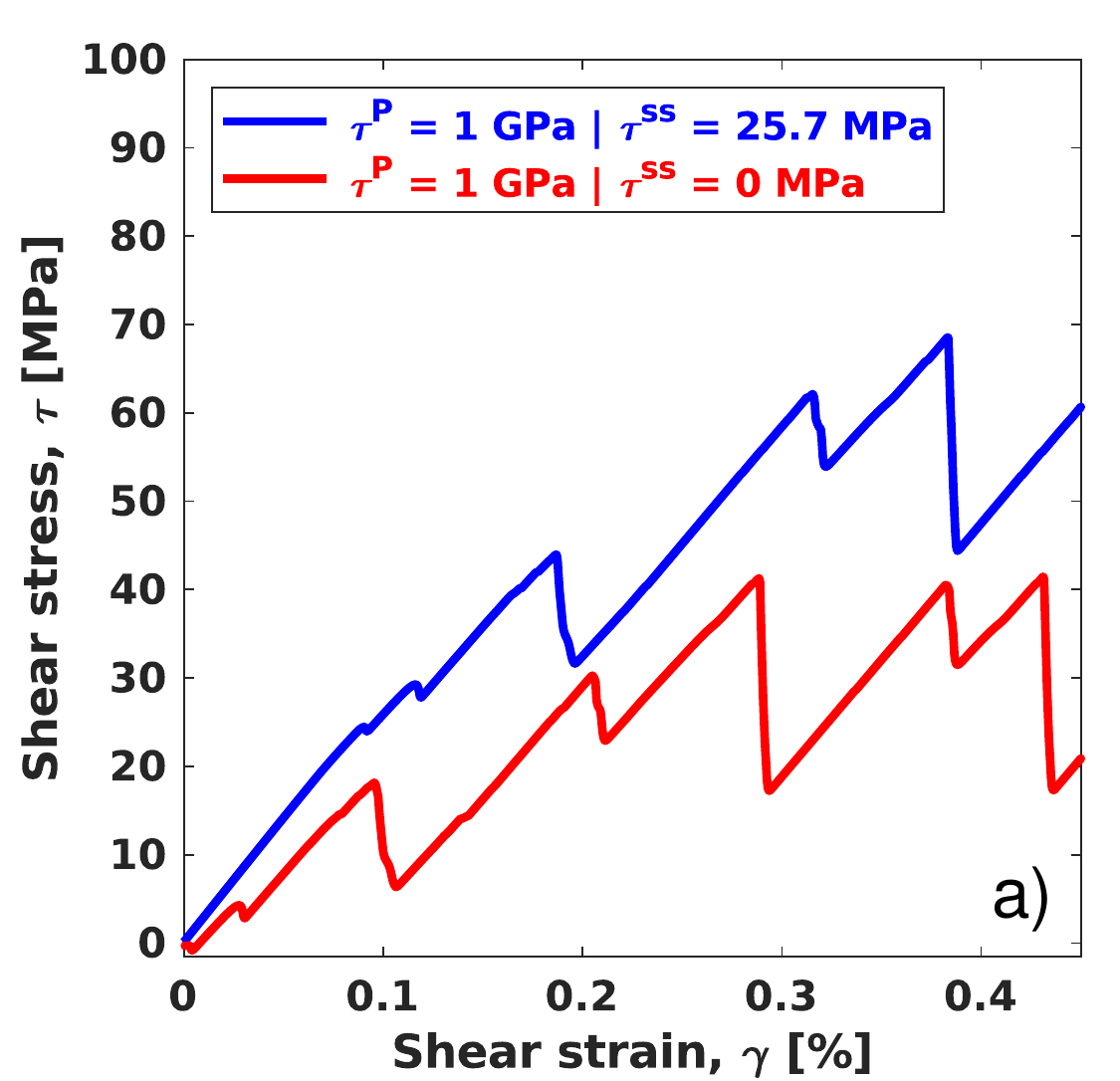}
\includegraphics[width=0.49\textwidth]{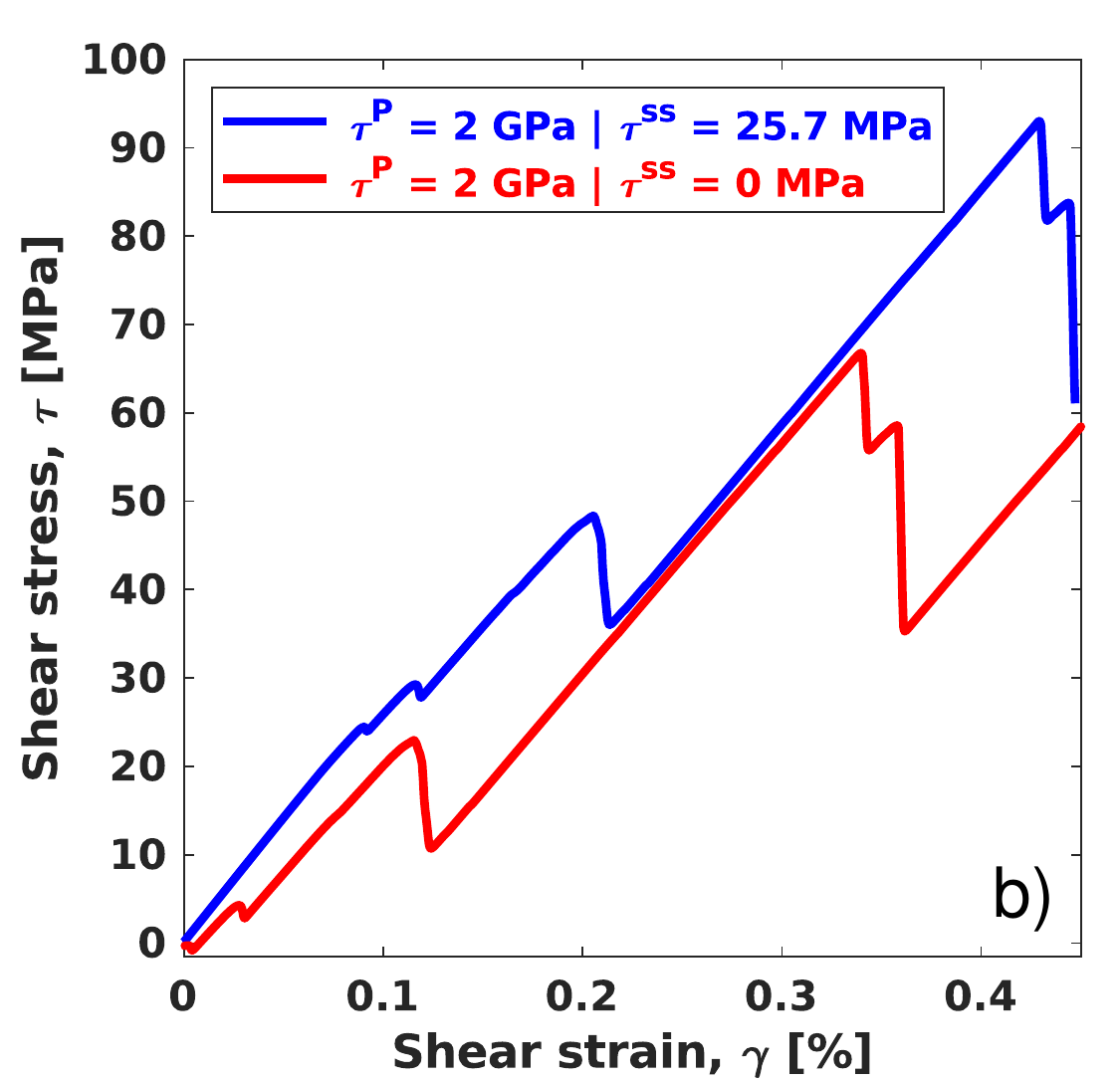}
\caption{Effect of solid solution hardening on the shear stress-shear strain curves obtained from DDD simulations of an initial edge dislocation. (a)  $\tau^P=1$ GPa. (b)  $\tau^P=2$ GPa.}  
\label{fig:stress_strain_curveSS}
\end{figure}

The shear stress-strain curves associated with the propagation of a dislocation through the simulation domain with and without including the effect of solid solution hardening are plotted in Figs. \ref{fig:stress_strain_curveSS}a and b when the $\tau^P$ = 1 and 2 GPa, respectively. Solid solution hardening increases the CRSS by an amount equal to $\tau^{ss}$ in both cases but the same peaks appear in the curves, regardless of whether solid solution is included. This indicates that solid solution hardening and precipitate hardening are additive and that the friction stress induced by solid solution hardening does not change the deformation mechanisms. The same conclusions were previously obtained for impenetrable precipitates \citep{Santos-Guemes2020}.

 \subsection{Comparison with experimental results}

\begin{figure}[t]
\centering
\includegraphics[width=0.75\textwidth]{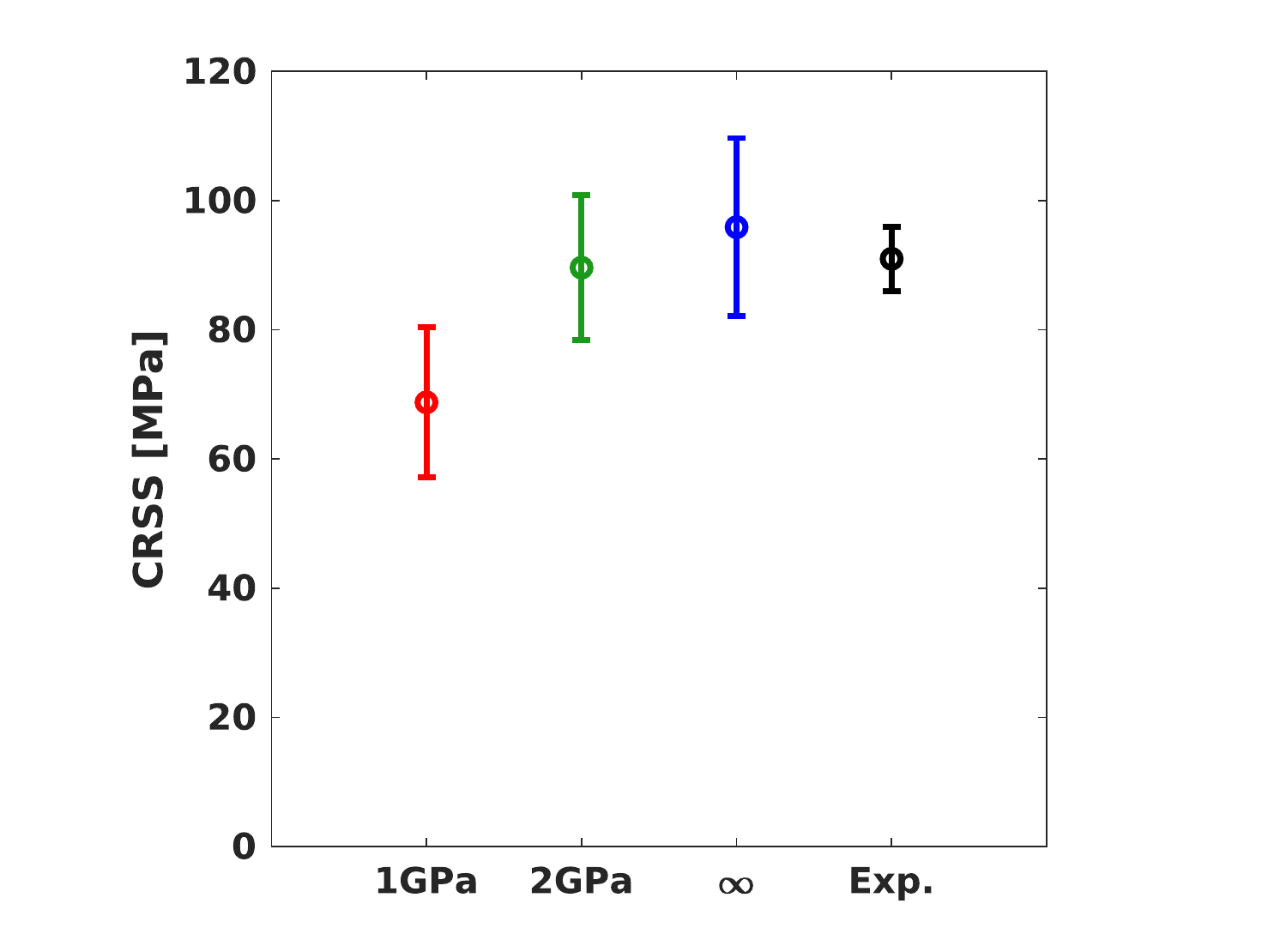}
\caption{DDD predictions of the CRSS of an Al-4 wt. {\%} Cu containing a random and homogeneous distribution of  1 vol. \% of $\theta''$ precipitates as a function of the  shear resistance of the precipitates $\tau^P$. The experimental results from \cite{Bellon2020} are included for comparison.}  
\label{fig:Stats}
\end{figure}

The average value and the corresponding standard deviation of the CRSS obtained from 10 simulations (5 with an initial edge dislocation and 5 with an initial screw dislocation) are plotted in Fig. \ref{fig:Stats} for different values of the threshold stress of the precipitates ($\tau^P$ = 1 GPa, 2 GPa, and $\infty$) together with the experimental results reported in \cite{Bellon2020}. The best agreement between the average experimental and simulated CRSS is obtained when $\tau^P$ = 2 GPa. The results of the simulations carried out with $\tau^P$ = 1 GPa underestimated the CRSS while the average CRSS of the simulations with impenetrable precipitates overestimated slightly the experimental results. Taking into account these results and the experimental evidence in Figs. \ref{fig:TEM_30h}b and c, it can be concluded that the $\theta''$ precipitates obtained in the Al - 4 wt. \% Cu alloy after 30 hours of ageing at $180^{\circ}$C were close to the transition between shearable/impenetrable precipitates, which is regarded as the optimum from the strengthening viewpoint as it increases strength while mitigating strain hardening \citep{ardell1985precipitation,Martin1998_PrecipitationHardening}.

\section{Discussion}
\label{sec:discussion}

The strengthening provided by a homogeneous distribution of $\theta''$ precipitates in an Al-Cu alloys was analyzed by means of DDD. Special care was taken to reproduce the experimental conditions and include the physical phenomena that influence the dislocation/precipitate interactions to ascertain whether the simulation was able to provide accurate predictions of the CRSS. To  this end, the shape, size distribution, orientation and volume fraction of the precipitates in the simulation domain were obtained from detailed transmission electron microscopy observations. In addition, the DDD simulations included the effect of elastic mismatch, coherency stresses and solution hardening, and the parameters that control these mechanisms were obtained from either first principle calculations or independent experiments. Finally, precipitate shearing was controlled by a threshold stress, $\tau^P$, that adopted three different values, 1 GPa, 2 GPa and $\infty$, the latter to represent impenetrable precipitates.  It was found that the average CRSS obtained from the simulations was in excellent agreement with the experimental one when $\tau^P$ = 2 GPa. Moreover, simulations showed a combination of precipitate shearing and formation of Orowan loops around the precipitates for this magnitude of the threshold stress, in agreement with the experimental observations. It should also be noted that the CRSS only increased slightly if the precipitates were assumed to be impenetrable by dislocations in the simulations and, thus, the actual magnitude of $\tau^P$ (insofar is $\ge$ 2 GPa) will not change significantly the CRSS. Of course, smaller $\tau^P$ will lead to a noticeable reduction of the CRSS, a shown in Fig. \ref{fig:Stats}.

It is important to notice that threshold stresses above 2 GPa will not change the initial CRSS of the alloy but the fact that the precipitates are shearable may have a large influence on the strain hardening behavior. The simulations in Figs.  \ref{fig:Snapshot_Edge04}a and \ref{fig:Snapshot_Screw02_2}a show that  that the precipitates  with $\tau^P$ = 1 GPa are sheared by dislocations and this phenomenon will be repeated every time a new dislocation interacts with the precipitates. Thus, they will not lead to strain hardening due to the accumulation of dislocations at the interface. If the threshold stress is higher, the precipitates may not be sheared by the first dislocation but they will be when the successive dislocations pile-up around the precipitates. This mechanism of shearing of precipitates has been recently reported in Mg alloys where pile-ups of basal dislocations are able to shear intermetallic precipitates which should have very large friction stresses due to complexity of their unit cells which lead to very long Burgers vectors \citep{EMP19, AL20, EAP20}. Successive shearing of the precipitates promotes slip localization in Mg alloys and is responsible for the limited strain hardening found in precipitation-hardened Mg alloys \citep{CCP19, AL20}.

 The results of the DDD simulations showed that the main strengthening mechanism in the Al-Cu alloy containing $\theta''$ precipitates is the interaction of the dislocation with the precipitates leading to shearing of the precipitate or to the formation of an Orowan loop, followed by the solution hardening contribution. On the contrary, the effect of coherency strains led to a reduction of the order of 1 MPa in the CRSS while the influence of the elastic mismatch between matrix and precipitate and cross-slip on the CRSS was negligible. These conclusions are in contrast with our previous DDD simulations \citep{Santos-Guemes2020} and experimental results \citep{Bellon2020} in the same Al-Cu alloy that was aged during 168 hours at 180$\degree$C to obtain an homogeneous dispersion of body center tetragonal $\theta'$ (Al$_2$Cu) precipitates. $\theta'$ precipitates also have a disk shape and grow parallel to the \{100\} planes of the $\alpha$-Al matrix  and the volume fraction of $\theta'$ precipitates after ageing was 1\% but their average diameter was  342 $\pm$ 47 nm, approximately one order of magnitude bigger than that of the $\theta''$ precipitates. These large precipitates could not be sheared by dislocations and their contribution to the strengthening of the alloy from the viewpoint of the Orowan mechanism was much smaller than that provided by a dispersion of smaller $\theta''$ precipitates. Nevertheless, the experimental values of the CRSS of the alloy after ageing during 30 and 168 hours at 180$\degree$C were very close (80 $\pm$ 7 MPa and 91 $\pm$ 5 MPa for the alloys containing $\theta'$ and $\theta''$ precipitates, respectively) and it should be noted that the amount of Cu in solid solution after ageing was also similar in both materials.

\begin{figure}[t!]
\centering
\includegraphics[width=0.6\textwidth]{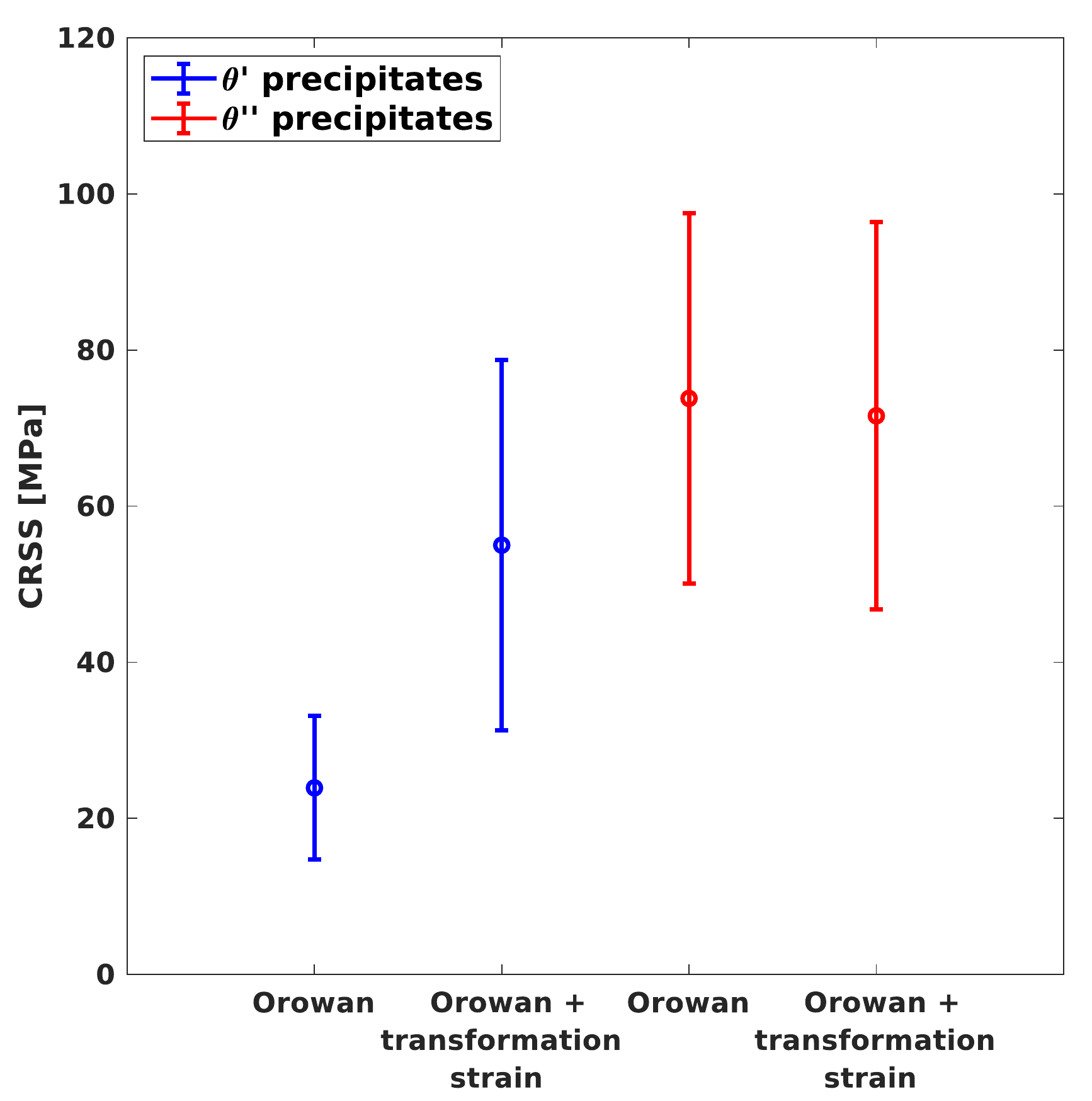}
\caption{DDD predictions of the maximum CRSS for  Al - 4 wt. \% Cu alloys containing 1 vol. \% of $\theta'$  and $\theta''$ impenetrable precipitates as a function of the dislocation/precipitate interaction mechanisms. The effect of the solid solution hardening is not included. The results for $\theta'$ precipitates can be found in \cite{Santos-Guemes2020}.}
\label{fig:Thetap_vs_Theta2p}
\end{figure}

In order to compare the mechanical response of alloys hardened by $\theta'$ and $\theta''$ precipitates, new DDD simulations were carried out assuming the same volume fraction of precipitates (1\%) in both cases and that all precipitates were impenetrable by dislocations.  49 precipitates were included in the simulation domains containing $\theta''$ and 12 precipitates in the simulation domains with $\theta'$.  The precipitate diameters were constant and equal to the average experimental value in each case. The contribution of solid solution hardening was not included in these results because it was the same for both alloys. The effect of the elastic mismatch was included in all cases although its contribution to the CRSS is negligible for both types of precipitates. The average values of the CRSS (obtained by averaging  the results of different DDD simulations)  for both alloys  are plotted in Fig. \ref{fig:Thetap_vs_Theta2p}. In the case of the $\theta''$ precipitates, the strengthening is mainly provided by the Orowan mechanism and the coherency strains associated with the formation of the $\theta''$ precipitates do not modify the CRSS. However, the Orowan contribution to the CRSS in the case of $\theta'$ precipitates is much smaller (slightly $>$ 20 MPa) and most of the strengthening (up to $\approx$ 55 MPa) was due to the presence of transformation strains associated with the formation of $\theta'$ precipitates. This transformation strain involves a relatively small volumetric component (about 6\%) and a homogeneous shear by an angle arctan(1/3) \citep{DW83, NM99}. The large shear stresses associated with this transformation strain hinder the propagation of the dislocation and enhance the CRSS by $\approx$ 35 MPa. They also promote dislocation cross-slip but the overall influence of this mechanism in CRSS was also negligible.

Thus, there are two ways to optimize the strength provided by precipitates in Al-Cu alloys. The first involves a homogeneous dispersion of small, coherent  $\theta''$ precipitates. $\theta''$ precipitates tend to nucleate homogeneously in the $\alpha$-Al matrix during high temperature ageing \citep{Liu2019}  and  the peak-aged condition is achieved when they attain  the critical size ($\approx$ 40 nm) which is associated with the transition from shearable to impenetrable.  On the contrary, $\theta'$ precipitates, impenetrable by dislocations, tend to nucleate heterogeneously on dislocations and grain boundaries at the initial stages of high temperature ageing \citep{Liu2017} and a homogeneous dispersion of $\theta'$ precipitates is only attained after much longer ageing times due to the nucleation of $\theta'$ on $\theta''$ precipitates that have grown up a critical size \citep{Liu2019}. In these conditions, a similar increase in the CRSS due to precipitation hardening can be achieved through the contributions of the stresses associated with the transformation strain and -to a minor extent- of the Orowan mechanism.

\section{Conclusions}
\label{sec:Conclusions}

The strength of  an Al - 4 wt. \% Cu alloy in the peak-aged condition, strengthened with a homogeneous dispersion of $\theta''$ precipitates, was analyzed by means of discrete dislocation dynamic simulations. The details of the microstructure (size distribution, shape, orientation and volume fraction of the precipitates) and of the dislocation/precipitate interactions (elastic mismatch, transformation strains, dislocation mobility and cross-slip probability, etc.) were obtained from experimental observations and atomistic simulations, respectively. The precipitates were assumed to be either impenetrable or shearable by the dislocations, the latter characterized by a threshold shear stress that has to be overcome to shear the precipitate.

The critical resolved shear stress calculated from the dislocation dynamics simulations was in very good agreement with the experimental results obtained from micropillar compression tests on single crystals oriented for single slip when the threshold stress for precipitate shearing was 2 GPa. The simulations showed a mixture of precipitate shearing and formation or Orowan loops around of the precipitate, also in agreement with the experimental observations. Simulations showed that the main strengthening mechanisms of the peak-aged Al-Cu alloy were provided by the precipitates (which have to be overcome by the dislocations through shearing or by forming an Orowan loop) and by the Cu atoms in solid solution. The effect of the elastic mismatch between matrix and precipitates, of the coherency stress and of dislocation cross-slip on the critical resolved shear stress was negligible.  Thus, the optimum strength of this precipitation-hardened alloy is attained by homogeneous distribution of $\theta''$ precipitates whose average size ($\approx$ 40 nm) is at the transition between precipitate shearing and looping.

In addition, the strengthening mechanisms of Al-Cu alloys containing homogeneous distributions of impenetrable $\theta''$ and $\theta'$ precipitates were analyzed using the same strategy. The critical resolved shear stress of the alloy with $\theta'$ precipitates was close to the one attained with $\theta''$ although the precipitates were much larger and the volume fraction in both cases was 1\%. It was found that the main strengthening mechanisms in the alloy with $\theta'$ precipitates were due to the transformation strains associated with the formation of $\theta'$ precipitates and the solid solution hardening, while the contribution of the formation of Orowan loops was limited due to the large spacing between precipitates. Overall, the dislocation dynamics strategy presented in this paper has demonstrated its capability to provide quantitative predictions of precipitate strengthening in metallic alloys.

\section{Acknowledgments}

This investigation was supported by the European Research Council under the European Union's Horizon 2020 research and innovation programme (Advanced Grant VIRMETAL, grant agreement No. 669141). RSG acknowledges the support from the Spanish Ministry of Education through the Fellowship FPU16/00770.



\end{document}